\documentclass[pdflatex,sn-nature]{sn-jnl}

\usepackage{graphicx}
\usepackage{multirow}
\usepackage{amsmath,amssymb,amsfonts}
\usepackage{booktabs}
\usepackage{array}
\usepackage{makecell}
\usepackage{xcolor}
\usepackage{soul}
\usepackage{textcomp}
\usepackage{url}
\usepackage[capitalize,noabbrev]{cleveref}
\crefname{figure}{Fig.}{Figs.}
\crefname{equation}{}{}
\crefname{section}{Section}{Sections}
\crefname{table}{Table}{Tables}

\begin{document}

\title[Ultra-Low-Energy Open-Circuit Fault Diagnosis for Three-Phase Inverters Using Neuromorphic Hardware]{Ultra-Low-Energy Open-Circuit Fault Diagnosis for Three-Phase Inverters}

\author[1]{\fnm{Xiaoyi} \sur{Lei}}
\author[1]{\fnm{Fanfu} \sur{Wu}}
\author*[1]{\fnm{Yunting} \sur{Liu}}\email{ypl5778@psu.edu}

\affil[1]{\orgname{The Pennsylvania State University}, \orgaddress{\state{PA}, \country{USA}}}

\abstract{Embedded fault diagnosis in three-phase inverters must satisfy the sub-watt power budget of converter control hardware, but conventional convolutional neural network (CNN)-based methods require dense multiply--accumulate operations and impose substantial inference energy. This work proposes an event-driven neuromorphic framework for energy-efficient open-circuit (OC) fault diagnosis. A CNN trained on current-vector trajectory matrices is converted into a spiking neural network (SNN) and evaluated using the NengoLoihi framework with Loihi-based neuromorphic energy estimation. By exploiting the sparse structure of trajectory matrices, the SNN activates computation only in informative regions instead of processing the full feature map densely. Experiments on a three-phase inverter platform show that the proposed method achieves 11~$\mu$J per diagnosis, corresponding to a 382$\times$ inference-energy reduction compared with a GPU-based CNN, while maintaining 100\% diagnostic accuracy. Robustness is further validated under unbalanced loading, current amplitude step changes, and injected measurement noise.}


\maketitle

\section{Introduction}

Data-driven methods offer rapid response and high diagnostic accuracy for power converter fault diagnosis and have been extensively investigated in recent years \cite{xia2019data,xia2021transferrable,cai2016data,gou2020online,wang2019fault,hang2023robust,luo2025enhanced,wang2025global}. However, these algorithms demand substantial energy, often orders of magnitude higher than traditional model-based approaches. Specifically, implementing a convolutional neural network (CNN) on embedded hardware, such as a field-programmable gate array (FPGA), often requires a power budget of tens of watts. This drastically exceeds the sub-watt auxiliary power budget of typical converter controllers, which is often approximately 0.9~W \cite{ti_powerest}.

Prior research shows that CNN accelerators consume 7.94--337~mJ per diagnosis \cite{chen2016eyeriss,chen2019eyeriss}, while commercial edge devices can reach 500~mJ \cite{hadidi2019characterizing}. At high sampling rates, these costs can scale to hundreds of watts, far exceeding the auxiliary power budget of most converters. This limits the adoption of data-driven diagnosis in practical inverter systems. In contrast, model-based methods require only 0.12~mJ \cite{rothenhagen2005performance}.

Spiking neural networks (SNNs) provide a promising solution by combining the low energy footprint of model-based methods with the high accuracy of data-driven models. SNNs are naturally suited for sparse trajectory analysis due to their event-driven nature; neuronal updates are triggered only by nonzero events. Even though sparse convolution can also reduce redundant arithmetic by processing only nonzero input elements \cite{graham2014spatially,graham2017submanifold}; however, the indices of nonzero elements must still be stored and transferred during computation, introducing additional data-movement overhead. Moreover, sparse convolution remains a clock-driven MAC-based computational framework, and the associated index-transfer requirement limits its compatibility with compute-in-memory (CIM) hardware.

Few works have explored the use of event-driven algorithms for energy reduction in power systems, including applications in power grid operation \cite{mahapatra2021power,yang2025reinforcement} and switching loss minimization in power converters \cite{liu2025combining,liu2025event,liu2026fcsmpc}. However, event-driven neuromorphic computing has not been sufficiently investigated for power converter fault diagnosis. To the best of authors' knowledge, this paper is the first literature that proposes an SNN-based OC fault diagnosis framework for three-phase inverters that preserves the diagnostic accuracy of CNNs while being compatible with neuromorphic hardware and CIM hardware. The proposed framework is implemented using the NengoLoihi framework, and its neuromorphic inference energy is estimated based on the Intel Loihi energy model \cite{davies2018loihi}. The proposed approach is compared with graphics processing unit (GPU)-based CNN methods \cite{lei2026online}, GPU-based SNN simulation \cite{arena2015mushroom}, and recent data-driven diagnosis methods. Experimental results show that the proposed method achieves 11~$\mu$J per diagnosis, corresponding to a 382$\times$ reduction relative to GPU-based CNNs, while maintaining 100\% accuracy. The method is further validated under balanced loading, extremely unbalanced loading, current amplitude step changes, and injected measurement noise, demonstrating the robustness and feasibility of the proposed method under various operating conditions.

\section{Results}

\subsection{Trajectory sparsity creates redundant CNN computation}

\begin{figure*}[t]
    \centering
    \includegraphics[width=\textwidth]{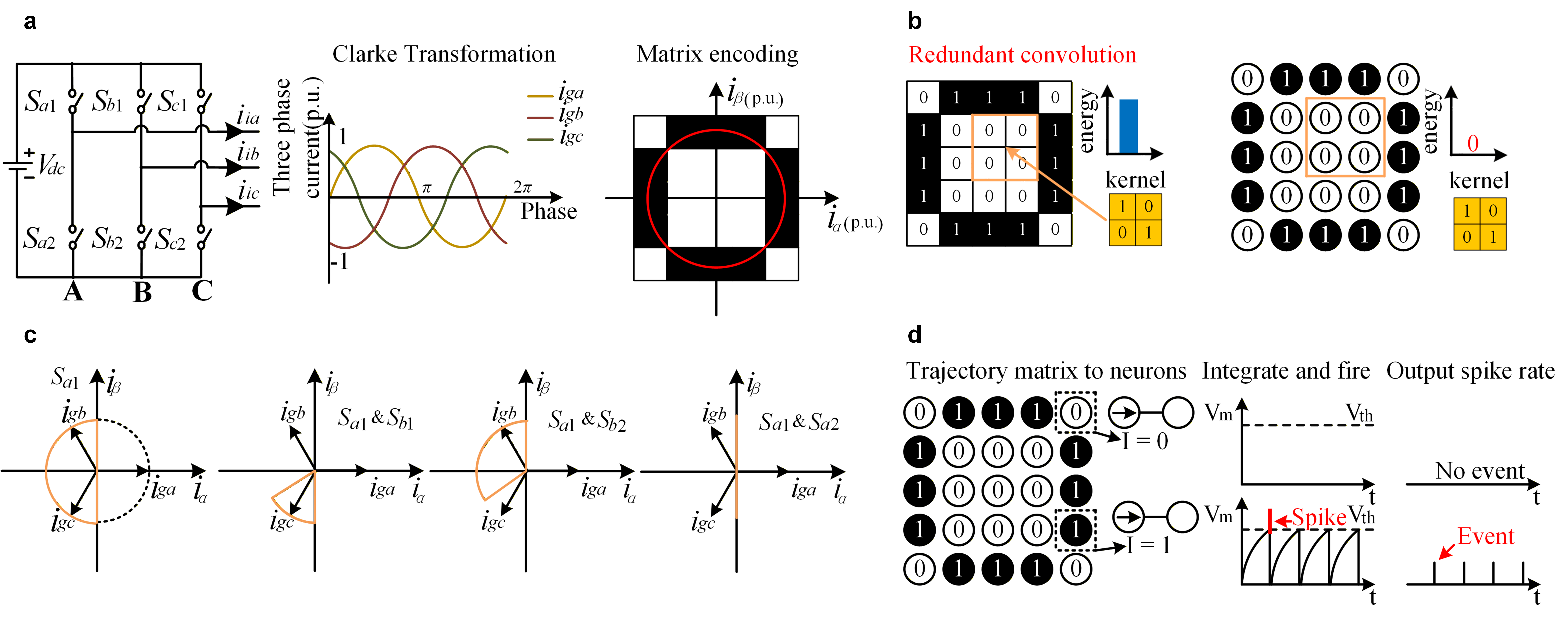}
    \caption{\textbf{Current-vector trajectory generation and spike encoding for OC fault diagnosis.} \textbf{a}, Three-phase inverter currents are transformed into the $\alpha$-$\beta$ domain and encoded as a binary trajectory matrix. \textbf{b}, Dense convolution over zero-valued trajectory regions introduces redundant computation and energy consumption. \textbf{c}, Representative current-vector trajectories generated by phase-A upper-switch OC fault, phase-A upper-switch and phase-B upper-switch OC fault, phase-A upper-switch and phase-B lower-switch OC fault, and phase-A upper-switch and phase-A lower-switch OC fault in the $\alpha$-$\beta$ domain. \textbf{d}, Trajectory-matrix elements are mapped to spiking neurons; zero-valued elements generate no spikes, whereas active elements produce event-driven spike trains.}
\label{fig:trajectory_spike}
\end{figure*}

Trajectory matrices are inherently sparse; for a $112 \times 112$ matrix dimension, only around 6\% of elements are nonzero. Despite this, the synchronous, clock-driven paradigm of CNNs requires kernels to be applied across the entire matrix. As shown in Fig.~\ref{fig:trajectory_spike}b, this results in massive computational redundancy, as the energy-intensive multiply-accumulate (MAC) operations are performed on zero-valued elements that contain no fault information.

The majority of MAC operations are redundant for OC fault diagnosis. For the four-layer CNN architecture proposed in \cite{lei2026online}, in the input convolutional layer, a total of 3.61 million MACs are generated to process the initial trajectory matrix. However, due to the physical nature of the open-circuit fault, the input data is extremely sparse. For a trajectory matrix with a resolution of $112 \times 112$, typically only 6\% of the elements are traversed by the current trajectory and marked as 1. Consequently, in the input layer, only a small fraction of the total MAC operations actually process informative features, while the remaining are computationally redundant.

The significant energy inefficiency caused by the mismatch between the sparsity of the trajectory matrix and the dense computation in CNN architectures can be mitigated by the event-driven computational paradigm of SNNs. Under healthy conditions, the blue region contains no activity; therefore, no computation is triggered in this region. In contrast, under the phase-A upper switch fault, the current trajectory is distorted, activating the blue region while the yellow region becomes inactive. As a result, the event-driven nature of SNNs enables computation to occur only in regions where current spikes are present, effectively exploiting the sparsity of the trajectory matrix.

\subsection{Neuromorphic implementation of SNN-based OC diagnosis}

\begin{figure*}[t]
    \centering
    \includegraphics[width=\textwidth]{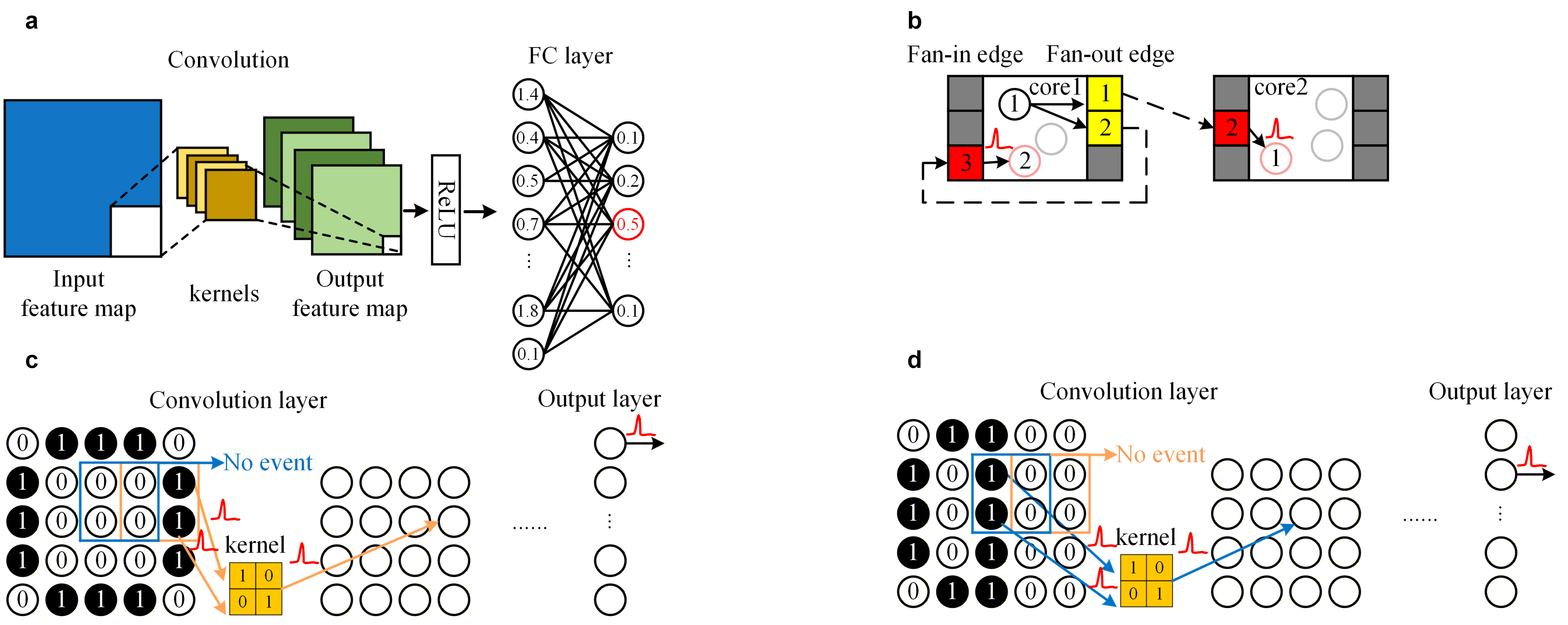}
    \caption{\textbf{CNN-to-SNN conversion and event-driven neuromorphic inference framework.} \textbf{a}, CNN architecture for trajectory-matrix-based OC fault diagnosis, consisting of convolutional layers and a fully connected output layer. \textbf{b}, Fan-in and fan-out connectivity used to map spike communication across Loihi neuromorphic cores \protect\cite{davies2018loihi}. \textbf{c}, SNN processing of a healthy-mode current-vector trajectory matrix. The blue region contains only zero-valued elements and therefore generates no events, while the yellow region contains trajectory information and triggers spike communication. \textbf{d}, SNN processing of a phase-A upper-switch OC fault current-vector trajectory matrix. The yellow region contains only zero-valued elements and therefore generates no events, while the blue region contains fault-related trajectory information and triggers spike communication.}
\label{fig:cnn_snn_loihi}
\end{figure*}

The CNN-based OC fault diagnosis framework is set as the baseline. CNN firstly extracts spatial features from the trajectory matrix using convolutional layers and then performs fault classification through a fully connected output layer. This structure provides the reference model for SNN conversion, as shown in Fig.~\ref{fig:cnn_snn_loihi}a.

After CNN training, the SNN can be obtained by replacing all ReLU activation functions with spiking neurons, as illustrated in Fig.~\ref{fig:cnn_snn_loihi}c,d. Only the input layer is visualized for simplicity. By mapping the trained CNN weights to synaptic weights in the SNN, connections between spiking neurons establish spike transmission pathways across layers. Different trajectory matrices therefore induce distinct spiking activity patterns in the output layer neurons.

The proposed framework follows the computation model of the Intel Loihi architecture, whose compute-in-memory (CIM) characteristics motivate the event-driven execution adopted in this work. As shown in Fig.~\ref{fig:cnn_snn_loihi}b, the circles represent spiking neurons. Each spiking neuron is associated with local processing and memory resources. As a result, the neuron state, synaptic parameters, and computation are physically organized close to the corresponding spiking neuron \cite{davies2018loihi}. Therefore, the proposed SNN-based diagnosis framework exploits sparsity from both the algorithm and hardware perspectives: zero-valued trajectory regions generate no spikes, while Loihi supports localized memory access and event-driven spike communication.

\subsection{Energy and accuracy under balanced loading}

The proposed method is first evaluated under balanced loading conditions using the three-phase inverter system parameters listed in \cref{tab:3}. Before accuracy evaluation, the firing-rate scaling factor and neuron firing threshold must be specified. In the Nengo Loihi workflow used in this study, the firing threshold is defined by the default Loihi spiking rectified-linear neuron model and is not treated as a manually tuned conversion parameter. Therefore, spiking activity is adjusted through the firing-rate scaling factor. Since prediction is obtained from the network output at the final time step, a higher firing rate enables the target neuron to emit more spikes, thereby improving diagnostic accuracy.

\begin{figure*}[t]
    \centering
    \includegraphics[width=0.95\textwidth]{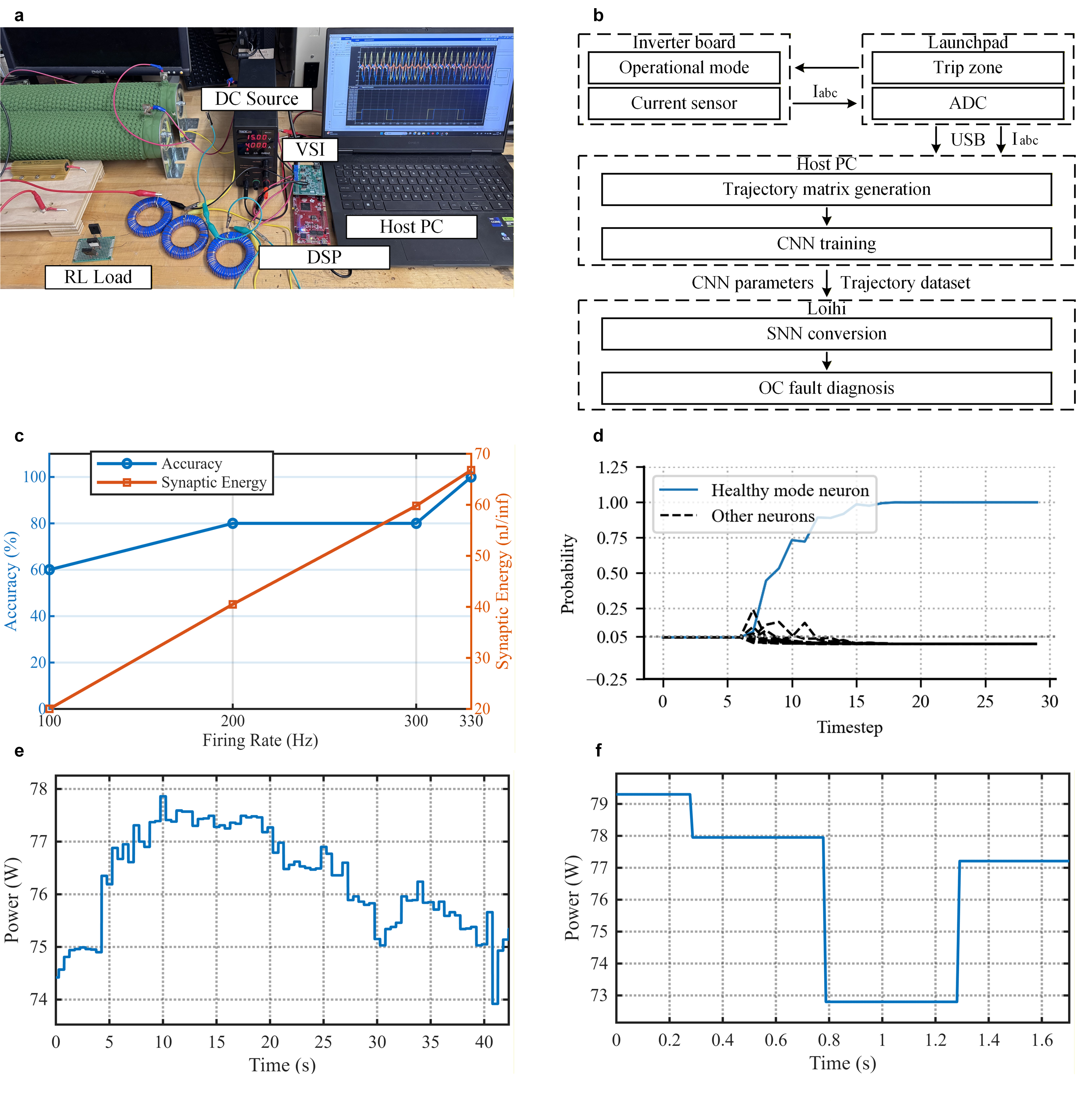}
    \caption{\textbf{Experimental validation workflow and energy-logging results.} \textbf{a}, Laboratory test bench for three-phase inverter OC fault diagnosis. \textbf{b}, Data acquisition, trajectory-matrix generation, CNN training, SNN conversion and NengoLoihi-based offline diagnosis workflow with Loihi energy estimation. \textbf{c}, Effect of firing-rate scaling factor on diagnostic accuracy and synaptic energy. \textbf{d}, Output-layer neuron probabilities under healthy-mode inference. \textbf{e}, GPU power logging profile during CNN training. \textbf{f}, GPU power profile during repeated offline CNN validation.}
\label{fig:experimental_workflow}
\end{figure*}

However, increasing the firing rate also leads to higher energy consumption due to the generation of more spikes. The trade-off between diagnostic accuracy and energy consumption is illustrated in Fig.~\ref{fig:experimental_workflow}c. The total energy consumption of the SNN is given by
\begin{equation}
    E_{total} = E_{synaptic} + E_{neuron},
    \label{eq:energy}
\end{equation}
where $E_{synaptic}$ denotes the energy consumed by spike generation and transmission, and $E_{neuron}$ represents the energy required for membrane voltage updates. In Loihi, membrane voltage updates are performed at every simulated time step regardless of synaptic activity, and the energy cost per neuron update is higher than that of individual synaptic events \cite{davies2018loihi}. As a result, neuron state updates introduce a non-negligible baseline energy consumption that is largely independent of input sparsity. In the considered implementation, the large input matrix size ($112 \times 112$) leads to a substantial number of active neurons, causing the neuron-update energy $E_{\text{neuron}}$ to reach the order of $\mu$J, which is approximately two orders of magnitude higher than the synaptic energy $E_{synaptic}$. As a result, the total energy per inference is dominated by $E_{neuron}$, as shown in \cref{tab:energy_estimation}, while $E_{synaptic}$ increases approximately linearly with the firing rate, as shown in Fig.~\ref{fig:experimental_workflow}c.

It is also observed that an accuracy plateau of approximately 80\% exists when the firing rate increases from 200~Hz to 300~Hz. This phenomenon is related to the spike-based prediction mechanism: when the firing rate is insufficient, spiking activity becomes sparse within the time window, and the target neuron may fail to fire at the final time step, resulting in incorrect prediction. To achieve higher accuracy, a sufficiently large scaling factor must be selected. However, since the maximum achievable accuracy of the SNN is bounded by that of the trained CNN, increasing the firing rate beyond this point only introduces redundant synaptic energy consumption. The selected firing-rate scaling factor is associated with the adopted operating condition, trajectory-matrix encoding, network architecture, and Loihi neuron model. If these conditions change significantly, the scaling factor should be re-calibrated using the same accuracy-based procedure.

With the selected scaling factor, the SNN achieves 100\% diagnostic accuracy on the 4,400-sample test set. The output neuron response under the healthy mode is used as an example to illustrate the prediction mechanism of the SNN in Fig.~\ref{fig:experimental_workflow}d. At the initial stage, no spikes are observed at the output layer. Since the output layer consists of 22 spiking neurons corresponding to 22 inverter operational modes, each neuron is associated with an equal probability, and the healthy-mode neuron initially accounts for $\frac{1}{22} \approx 0.045$. As the trajectory matrix is presented to the input layer, spiking neurons are gradually activated, generating current spikes that propagate through the network toward the output layer. By mapping the trained CNN weights to synaptic weights in the SNN, trajectory matrices with circular patterns corresponding to the healthy mode stimulate the healthy-mode neuron more strongly, resulting in higher spiking activity in that neuron. As the simulation proceeds over time steps, the healthy-mode neuron continues to fire spikes, while the other neurons remain largely inactive. Consequently, the probability associated with the healthy mode becomes dominant by the end of the simulation window.

Table~\ref{tab:energy_estimation} summarizes the energy comparison. Training energy and testing energy denote the total energy consumed during training and one round of testing, respectively. The energy per inference is computed by dividing the testing energy by the number of inferences (4,400). Among the GPU-based methods, the KAN achieves the lowest energy per inference (\(1.2 \times 10^{-3}\)~J), as it contains the smallest number of learnable parameters. Consequently, each inference requires fewer MAC operations, resulting in the lowest energy consumption among the GPU-based approaches. For other methods (WT + CNN, TCN, and A-MICNN), which are based on 1-D CNN architectures, the number of learnable parameters is also relatively small, compared to 2-D CNN-based models. As a result, these methods exhibit lower energy consumption per inference compared to 2-D CNN-based models.

In contrast, GPU-based SNN models consume significantly more energy. This is because their computation emulates spiking neuron dynamics using integrate-and-fire mechanisms, where each inference is divided into multiple state-variable calculations (membrane voltage, current spike transmission, etc.) over multiple simulated time steps \cite{ottati2024efficient}. Consequently, the number of MAC operations is substantially increased. Specifically, the mushroom body (MB) model achieves an accuracy of 78.32\% with an energy consumption of 0.17~J per inference. In contrast, NengoLoihi achieves energy consumption of \({1.1 \times 10^{-5}}\)~J per inference, representing a 382$\times$ reduction relative to the GPU-based CNN and an average 20,000$\times$ reduction relative to GPU-based SNN simulators, without any loss of diagnostic accuracy. Compared with all GPU-based non-SNN data-driven methods in \cref{tab:energy_estimation}, an average 239$\times$ reduction is achieved, while maintaining higher or comparable diagnostic accuracy.

\begin{table*}[t]
\centering
\caption{Comparison study in energy consumption and diagnostic accuracy under balanced loading}
\label{tab:energy_estimation}
\footnotesize
\setlength{\tabcolsep}{3pt}
\resizebox{\textwidth}{!}{%
\begin{tabular}{lcccc}
\toprule
\textbf{Methodology} & \textbf{Accuracy (\%)} &
\textbf{Training Energy (J)} &
\textbf{Testing Energy (J)} &
\textbf{Energy per Inference (J)} \\
\midrule

NengoLoihi (CNN-to-SNN) & 100\% & 2844.53 & \(\mathbf{0.048}\) & \(\mathbf{1.1 \times 10^{-5}}\) \\

Laptop NVIDIA 4060 (CNN offline) \cite{lei2026online} & 100\% &
2844.53 & 18.48 & \(4.2 \times 10^{-3}\) \\

Laptop NVIDIA 4060 (CNN online) \cite{lei2026online} & 100\% &
2844.53 & 24.64  & \(5.6 \times 10^{-3}\) \\

Laptop NVIDIA 4060 (CNN to SNN) \cite{diehl2015fast} & 100\% &
2844.53 & 1364 & 0.31 \\

Laptop NVIDIA 4060 (MB) \cite{arena2015mushroom} & 78.32\% &
2952.11 & 748 & 0.17 \\

% Laptop NVIDIA 4060 (SS) \cite{zenke2018superspike} & 100\% &
% 2341.71 & 792 & 0.18 \\

Laptop NVIDIA 4060 (WT + CNN)\,\cite{hang2023robust}
& 100\% & 1363.3 & 8.73 & \(2 \times 10^{-3}\) \\

Laptop NVIDIA 4060 (KAN)\,\cite{wang2025global}
& 99.25\% & 894.89 & 5.15 & \(1.2 \times 10^{-3}\) \\

Laptop NVIDIA 4060 (TCN)\,\cite{lu2025novel}
& 98.48\% & 831.93 & 5.74 & \(1.3 \times 10^{-3}\) \\

Laptop NVIDIA 4060 (A-MICNN)\,\cite{mahmoud2025robust}
& 99.74\% & 1537.18 & 6.77 & \(1.5 \times 10^{-3}\) \\

\bottomrule
\end{tabular}
}

\end{table*}

\subsection{Generalization to unbalanced loading}

The proposed method is also evaluated under extremely unbalanced loading conditions. The key parameters of the unbalanced load are listed in \cref{tab:un_para}. Under this condition, a PCUR of 88\% is observed. The firing-rate scaling factor is fixed at 330~Hz for consistency. The comparison results are summarized in \cref{tab:energy_estimation_UB}.

Compared to the results under balanced loading conditions, an average accuracy reduction of 5\% and an average increase of 18\% in training energy are observed. This is attributed to the increased convergence complexity introduced by the unbalanced operating condition. Moreover, GPU-based SNN simulators \cite{diehl2015fast,arena2015mushroom} still exhibit significantly higher energy consumption per inference, approximately 100$\times$ larger than other GPU-based methods. This increase is primarily due to the larger number of MAC operations required, as discussed previously.

Among all methods in the comparison study, the proposed method achieves the lowest energy per inference of \(1.1 \times 10^{-5}\)~J. Notably, this value remains unchanged compared to the balanced loading case. This is because the dominant component of energy consumption is $E_{neuron}$, which is approximately $160\times$ larger than $E_{synaptic}$. Since $E_{neuron}$ is determined by the SNN architecture and remains independent of input conditions, the overall energy per inference remains effectively constant as long as the network structure is unchanged.

\begin{table*}[t]
\centering
\caption{Comparison study in energy consumption and diagnostic accuracy under unbalanced loading}
\label{tab:energy_estimation_UB}
\footnotesize
\setlength{\tabcolsep}{3pt}
\resizebox{\textwidth}{!}{%
\begin{tabular}{lcccc}
\toprule
\textbf{Methodology} & \textbf{Accuracy (\%)} &
\textbf{Training Energy (J)} &
\textbf{Testing Energy (J)} &
\textbf{Energy per Inference (J)} \\
\midrule

NengoLoihi (CNN-to-SNN) & 100\% & 3238.78 & \(\mathbf{0.048}\) & \(\mathbf{1.1 \times 10^{-5}}\) \\

Laptop NVIDIA 4060 (CNN offline) \cite{lei2026online} & 100\% &
3238.78 & 13.42 & \(3 \times 10^{-3}\) \\

Laptop NVIDIA 4060 (CNN online) \cite{lei2026online} & 100\% &
3238.78 & 23.73  & \(5.2 \times 10^{-3}\) \\

Laptop NVIDIA 4060 (CNN to SNN) \cite{diehl2015fast} & 100\% &
3238.78 & 1682.46 & 0.37 \\

Laptop NVIDIA 4060 (MB) \cite{arena2015mushroom} & 65.47\% &
2844.71 & 784.53 & 0.18 \\

% Laptop NVIDIA 4060 (SS) \cite{zenke2018superspike} & 100\% &
% 2517.63 & 840.26 & 0.19 \\

Laptop NVIDIA 4060 (WT + CNN)\,\cite{hang2023robust}
& 93.11\% & 1929.75 & 12.09 & \(2.7 \times 10^{-3}\) \\

Laptop NVIDIA 4060 (KAN)\,\cite{wang2025global}
& 95.2\% & 944.26 & 8.27 & \(1.9 \times 10^{-3}\) \\

Laptop NVIDIA 4060 (TCN)\,\cite{lu2025novel}
& 93.18\% & 946.18 & 6.03 & \(1.4 \times 10^{-3}\) \\

Laptop NVIDIA 4060 (A-MICNN)\,\cite{mahmoud2025robust}
& 99.12\% & 2065.48 & 11.67 & \(2.7 \times 10^{-3}\) \\

\bottomrule
\end{tabular}
}

\end{table*}

\subsection{Robustness under current step changes and measurement noise}

\begin{figure*}[t]
    \centering
    \includegraphics[width=0.95\textwidth]{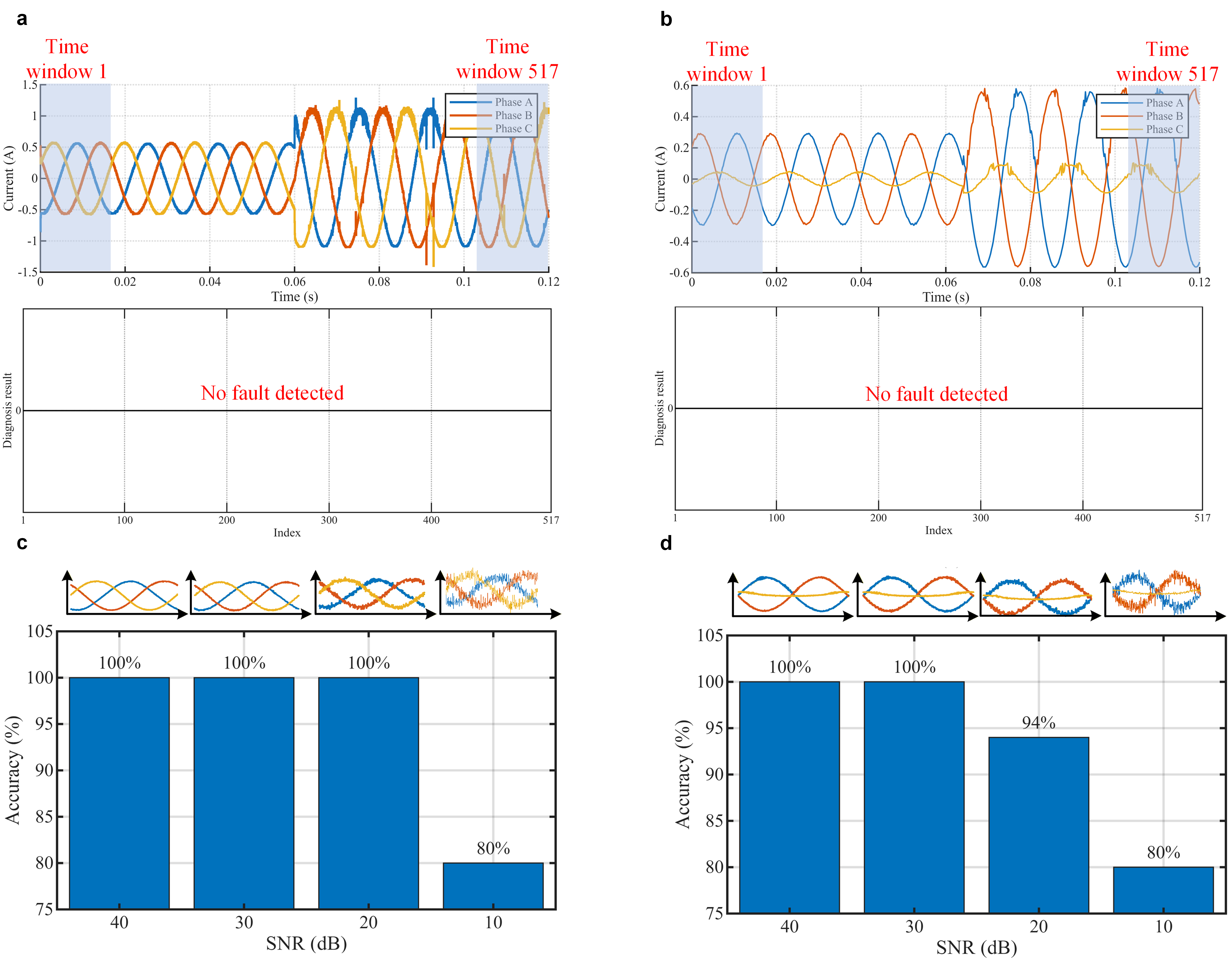}
    \caption{\textbf{Robustness against current step changes and injected measurement noise.} \textbf{a}, Three-phase current waveform and diagnosis result under balanced loading with current amplitude step change. \textbf{b}, Three-phase current waveform and diagnosis result under unbalanced loading with current amplitude step change. \textbf{c}, Offline diagnostic accuracy under SNR level of 40dB, 30dB, 20dB and 10dB for balanced loading. \textbf{d}, Offline diagnostic accuracy under SNR level of 40dB, 30dB, 20dB and 10dB for unbalanced loading.}
\label{fig:robustness}
\end{figure*}

The proposed method is further validated under current amplitude step-change scenarios to evaluate its robustness under dynamic operating conditions. Both balanced loading (Fig.~\ref{fig:robustness}a) and unbalanced loading (Fig.~\ref{fig:robustness}b) cases are considered. In both cases, the current amplitude changes from 0.5~p.u. to 1~p.u. to emulate a sudden load variation. A total duration of 0.12~s three-phase current is applied for evaluation.

Since the present implementation is evaluated using the NengoLoihi framework in an offline workflow, the three-phase current is first processed offline to generate current-vector trajectory matrices, which are then sent into the SNN for sequential diagnosis. During offline three-phase current processing, a sliding window approach is adopted for trajectory matrix generation. The window length is set to one fundamental cycle of the three-phase current, and the window advances by one sampling point at each step. As a result, each trajectory matrix corresponds to one diagnosis. The horizontal axis of the diagnosis results therefore represents the sliding window index rather than time.

With a sampling frequency of 5~kHz, the 0.12~s current signal corresponds to 600 sampling points. In the adopted implementation, an 83-point sliding window generates 517 diagnosis instances. Consequently, the sliding window index ranges from 1 to 517.

The diagnosis process can be divided into three stages: diagnosis at 0.5 p.u. current amplitude, the current-amplitude step transient, and diagnosis at 1 p.u. current amplitude. In the first and third stages, the current-vector trajectories are generated under healthy operation and therefore exhibit a circular pattern, so no fault is detected. During the amplitude step change transient, the corresponding trajectories are labeled as healthy during training; consequently, the SNN does not trigger a fault diagnosis during this interval as well, as shown in Fig.~\ref{fig:robustness}a,b. These results demonstrate that the proposed method is robust to current-amplitude step changes.

It should be noted that the apparent THD increase after the loading step change is mainly caused by the limited resolution of the 8-bit data transmission. At low current amplitude, low-amplitude harmonic components can be suppressed by quantization and are not clearly reflected in the reconstructed waveform. After the current amplitude increases, these harmonic components become more visible under the same 8-bit resolution, resulting in a higher apparent THD in the plotted waveform.

The robustness of the proposed method is evaluated under manually injected white Gaussian noise. The noise level is characterized by the signal-to-noise ratio (SNR), with values of 40~dB, 30~dB, 20~dB, and 10~dB applied to the three-phase currents. For each SNR level, 200 cycles of three-phase current are randomly sampled from each operational mode, resulting in a total of 4,400 current-vector trajectory matrices as the test dataset.

The proposed method is evaluated under both balanced and unbalanced loading conditions. The SNN models are trained using noise-free three-phase current data. Only offline validation is conducted, as real-time deployment on Loihi is currently infeasible. The firing-rate scaling factor is fixed at 330~Hz, consistent with the value listed in \cref{tab:cnn_architecture}. Robustness is quantified by the offline diagnostic accuracy, as shown in Fig.~\ref{fig:robustness}c,d.

One cycle of the three-phase current waveform is included at the top of the figure to illustrate the distortion introduced by noise. Under balanced loading conditions, the proposed method maintains 100\% diagnostic accuracy down to 20~dB SNR. At 10~dB SNR, severe waveform distortion leads to a reduction in diagnostic accuracy to 80\%.

Under unbalanced loading conditions, 100\% diagnostic accuracy is achieved at both 40~dB and 30~dB SNR. A 6\% reduction in accuracy is observed at 20~dB SNR, and at 10~dB SNR, the accuracy decreases to 80\%. The accuracy reduction under simultaneous unbalanced loading and noise is caused by the combined effect of trajectory shrinking and noise-induced distortion. Under extremely unbalanced loading, the small-amplitude phase current shrinks the trajectory along the corresponding projection direction in the $\alpha$-$\beta$ domain. As a result, the trajectory area decreases, and noisy data points are more likely to interfere with each other, causing greater trajectory distortion and partial overlap among similar OC fault patterns at low SNR.

These offline validation results demonstrate the robustness of the proposed method against white Gaussian noise. This noise immunity also supports the feasibility of future embedded deployment, where practical power converter systems are subject to measurement noise, sampling errors, and harmonic disturbances.

\subsection{Embedded implementation implications}

\begin{table*}[t]
\centering
\caption{Qualitative Comparison of Hardware Platforms for Embedded Implementation}
\label{tab:hardware_comparison}
\footnotesize
\setlength{\tabcolsep}{4pt}
\begin{tabular}{lcccc}
\toprule
\textbf{Platform} & \textbf{Cost} & \textbf{Performance} & \textbf{Ease of Use} & \makecell{\textbf{Embedded}\\\textbf{Availability}} \\
\midrule
Intel Loihi & N/A & High & Low & N/A \\
FPGA & Medium & Medium -- High & Low & High \\
STM32 MCU & Low & Low & High & High \\
RK3588 SoC & Medium & Medium & Medium & High \\
GPU & High & High & High & Low \\
\bottomrule
\end{tabular}
\end{table*}

The proposed method also shows potential for future embedded implementation. A qualitative comparison among candidate platforms is provided in \cref{tab:hardware_comparison}. Conventional embedded platforms either have limited computational resources, such as STM32-class MCUs, or still rely on Von Neumann architectures, such as RK3588 SoCs and embedded GPUs. Although embedded GPUs provide strong inference capability, their cost and power consumption deviate from the low-power motivation of this study. In contrast, FPGAs support flexible distributed memory using on-chip block RAM (BRAM), allowing SNN weights and neuron states to be stored close to the computational logic. This enables FPGA implementations to emulate the compute-in-memory characteristics of neuromorphic hardware while remaining more practical for converter-controller integration. Therefore, the proposed SNN-based method provides a promising pathway toward low-power monitoring in practical power electronic systems under sub-watt constraints.

\section{Discussion}

This work demonstrates that SNN-based OC fault diagnosis can substantially reduce inference energy without sacrificing diagnostic accuracy. Implemented on Intel NengoLoihi framework, the proposed method achieves an inference energy of $11~\mu$J per diagnosis, corresponding to a $382\times$ reduction compared with a GPU-based CNN implementation while maintaining 100\% diagnostic accuracy under balanced loading. The method also preserves robust diagnosis under extremely unbalanced loading, current amplitude step changes, and injected measurement noise.

The energy reduction is mainly enabled by matching the sparse structure of current-vector trajectory matrices with event-driven computation. For a $112 \times 112$ trajectory matrix under healthy operation, only a small fraction of elements contain trajectory information, while the remaining regions are non-informative. In conventional CNN implementations, dense convolution is still applied across the full matrix, causing energy consumption to scale with the total number of convolution and memory-access operations. In contrast, the SNN triggers spike communication only when informative trajectory elements are active, thereby avoiding redundant computation in zero-valued regions.

The CIM characteristic of neuromorphic hardware further supports this energy reduction by reducing data movement. In conventional Von Neumann hardware, input data and model parameters must be repeatedly transferred between global memory and processing units. In Loihi, neuron states and synaptic weights are stored in distributed on-chip memory close to the corresponding spiking neurons, allowing localized event-driven computation.

Despite these advantages, the Loihi-based implementation is not yet an optimized embedded realization. Although SNNs ideally consume energy primarily during spike generation, Loihi also incurs energy consumption from neuron state updates. In this implementation, $E_{neuron}$ dominates the total energy consumption, limiting the achievable energy saving compared with the theoretical expectation of purely event-driven computation.

Inference latency is another open issue. The time steps in Loihi correspond to simulated time rather than directly measured physical time, making it difficult to quantify absolute real-time detection delay in the current offline workflow. Future work will therefore focus on direct board-level power measurement and integration of the proposed neuromorphic diagnosis framework into embedded inverter controllers.

Together, these results indicate that neuromorphic hardware provides a promising pathway for implementing data-driven OC fault diagnosis under sub-watt constraints. The qualitative platform comparison further suggests that FPGA-based implementations may provide a practical route toward embedded inverter-controller integration by supporting distributed memory and localized computation.

\section{Methods}

\subsection{Open-circuit fault modes and trajectory matrix generation}

This paper focuses on OC fault diagnosis in a three-phase voltage source inverter (VSI). The 22 operational modes include one healthy mode, 6 single-switch OC faults, and 15 double-switch OC fault combinations among the six inverter switches, i.e., \(1+6+C_6^2=22\). The corresponding operational modes and labels are summarized in \cref{tab:operation_modes}.

The phase currents are transformed into the $\alpha$-$\beta$ frame using the Clarke transformation. As visualized in Fig.~\ref{fig:trajectory_spike}c, each operational mode generates a distinct current-vector trajectory geometry. These trajectories are encoded into binary trajectory matrices following \cite{lei2026online} and \cref{eq:encode}.

\begin{table*}[t]
\centering
\caption{Inverter operational modes and corresponding labels}
\label{tab:operation_modes}
\small
\renewcommand{\arraystretch}{1.08}
\setlength{\tabcolsep}{8pt}

\begin{tabular*}{\textwidth}{@{\extracolsep{\fill}} l c l c @{}}
\toprule
\textbf{Operational mode} & \textbf{Label}
& \textbf{Operational mode} & \textbf{Label} \\
\midrule
Healthy mode & \([0,0,0,0,0,0]\)
& \(S_{a1}\,\&\,S_{c2}\) fault & \([1,0,0,0,0,1]\) \\

\(S_{a1}\) fault & \([1,0,0,0,0,0]\)
& \(S_{a2}\,\&\,S_{b1}\) fault & \([0,1,1,0,0,0]\) \\

\(S_{a2}\) fault & \([0,1,0,0,0,0]\)
& \(S_{a2}\,\&\,S_{b2}\) fault & \([0,1,0,1,0,0]\) \\

\(S_{b1}\) fault & \([0,0,1,0,0,0]\)
& \(S_{a2}\,\&\,S_{c1}\) fault & \([0,1,0,0,1,0]\) \\

\(S_{b2}\) fault & \([0,0,0,1,0,0]\)
& \(S_{a2}\,\&\,S_{c2}\) fault & \([0,1,0,0,0,1]\) \\

\(S_{c1}\) fault & \([0,0,0,0,1,0]\)
& \(S_{b1}\,\&\,S_{b2}\) fault & \([0,0,1,1,0,0]\) \\

\(S_{c2}\) fault & \([0,0,0,0,0,1]\)
& \(S_{b1}\,\&\,S_{c1}\) fault & \([0,0,1,0,1,0]\) \\

\addlinespace[2pt]

\(S_{a1}\,\&\,S_{a2}\) fault & \([1,1,0,0,0,0]\)
& \(S_{b1}\,\&\,S_{c2}\) fault & \([0,0,1,0,0,1]\) \\

\(S_{a1}\,\&\,S_{b1}\) fault & \([1,0,1,0,0,0]\)
& \(S_{b2}\,\&\,S_{c1}\) fault & \([0,0,0,1,1,0]\) \\

\(S_{a1}\,\&\,S_{b2}\) fault & \([1,0,0,1,0,0]\)
& \(S_{b2}\,\&\,S_{c2}\) fault & \([0,0,0,1,0,1]\) \\

\(S_{a1}\,\&\,S_{c1}\) fault & \([1,0,0,0,1,0]\)
& \(S_{c1}\,\&\,S_{c2}\) fault & \([0,0,0,0,1,1]\) \\
\bottomrule
\end{tabular*}
\end{table*}

In this framework, phase currents are mapped to a $D \times D$ binary matrix $\mathbf{X}$ in the $\alpha$-$\beta$ plane using \cref{eq:encode}:
\begin{equation}
\label{eq:encode}
\mathbf{X}(m,n) =
\begin{cases}
    1, & m,n = \lfloor I_{\alpha,\beta}(t_i) \cdot \tfrac{D}{4} + \tfrac{D}{2} \rfloor \\
    0, & \text{otherwise}
\end{cases}
\end{equation}

\subsection{CNN formulation and computational overhead}

The CNN baseline extracts spatial features from trajectory matrices using convolutional layers. The final feature maps are flattened and passed to a fully connected layer, followed by Softmax classification across the 22 operational modes. The predicted fault mode is selected as the class with the highest probability, as illustrated in Fig.~\ref{fig:cnn_snn_loihi}a.

As shown in Fig.~\ref{fig:cnn_snn_loihi}a, the entire input matrix is required to be swept to identify trajectory geometrical distortions. While this architecture is highly effective at capturing the non-linear relationship between current trajectory shapes and specific fault modes, the dense nature of these operations across the entire input matrix creates a constant computational burden.

The computational complexity of the CNN is quantified by the number of multiply-accumulate (MAC) operations required per inference. For a standard architecture, these operations are defined as \cref{eq:MAC_conv} and \cref{eq:MAC_fc}:
\begin{equation}
MAC_{\text{conv}} =
(K_w \cdot K_h \cdot C_{\text{in}})
(H_{\text{out}} \cdot W_{\text{out}} \cdot C_{\text{out}})
\label{eq:MAC_conv}
\end{equation}
\begin{equation}
MAC_{\text{fc}} =
N_{\text{in}} \cdot N_{\text{out}}
\label{eq:MAC_fc}
\end{equation}
where $(K_w, K_h)$ denotes the kernel dimensions, $C_{in}$ and $C_{out}$ are the numbers of input and output channels, respectively, and $(H_{out}, W_{out})$ represents the spatial dimensions of the output feature map. For the fully connected (FC) layers, $N_{in}$ and $N_{out}$ denote the number of input and output neurons.

\subsection{CNN-to-SNN conversion and Loihi deployment}

A rate-based learning strategy was first proposed in \cite{diehl2015fast} and is adopted in this paper. To ensure that the outputs of the convolutional layers in the CNN can be interpreted as current spikes in the converted SNN, three compatibility constraints are imposed during CNN training:

\begin{enumerate}
    \item \textbf{Bias terms are fixed to zero:}
    Since the convolution outputs are interpreted as current spikes, a nonzero bias term would imply that spikes are generated even when no pre-synaptic spiking activity is present, which conflicts with the operating mechanism of SNNs. Therefore, all bias terms are fixed to zero during CNN training to ensure compatibility with CNN-to-SNN conversion.

    \item \textbf{ReLU is used to ensure nonnegative activations:}
    The ReLU activation function preserves a linear relationship between the input and output when the input is positive. Since spiking neurons also exhibit an approximately linear relationship between received current spikes and emitted spikes, all activation functions are set to ReLU during CNN training and are later replaced by spiking neurons during SNN conversion.

    \item \textbf{Pooling layers are omitted:}
    Pooling layers in CNNs are commonly used to reduce the dimensionality of feature maps through max-pooling or average-pooling operations over local regions. However, such operations are difficult to interpret in the temporal spike domain, and no direct counterpart exists in the converted SNN. Therefore, all pooling layers are omitted from the CNN architecture to maintain compatibility with SNN implementation.
\end{enumerate}

The behavior of spiking neurons can be modeled as leaky capacitors that accumulate charge and gradually discharge over time, while the trajectory matrix can be modeled as constant current sources that charge the corresponding spiking neurons. In the input layer, each element $x_{ij}$ of the trajectory matrix is assigned to a dedicated spiking neuron. The element value is converted into a driving current $I_{ij}$, such that larger matrix values produce stronger input currents. Each neuron follows an integrate-and-fire mechanism expressed in \cref{eq:LIF_in}: the membrane voltage $V_m$ increases under current injection until it reaches the threshold $V_{th}$, at which point the neuron emits a spike and the membrane voltage is reset to $V_{reset}$. If the input current is zero, the membrane voltage remains at the resting potential $V_{rest}$ and no spikes are generated. For simplicity, $V_{reset}$ is set equal to $V_{rest}$. This encoding process is illustrated in Fig.~\ref{fig:trajectory_spike}d.

\begin{equation}
\tau_m \frac{dV_m(t)}{dt} = -\left( V_m(t) - V_{rest} \right) + R \cdot I(t),
\label{eq:LIF_in}
\end{equation}

\noindent where $\tau_m=RC$ is the membrane time constant, which dictates the leakage rate of the capacitor.

In subsequent layers, neurons receive discrete current spikes generated by the preceding layer. The current spikes are marked as events in Fig.~\ref{fig:trajectory_spike}d. The spike train is modeled as a sequence of impulses using the Dirac delta function. $V_m$ increases stepwise upon receiving a spike and decays exponentially during the intervals between spikes. This behavior is described by:

\begin{equation}
\tau_m \frac{dV_m(t)}{dt} = -\left( V_m(t) - V_{rest} \right) + R \sum_{j} \omega \cdot \delta(t - t_j),
\label{eq:LIF_2}
\end{equation}

\noindent where $\delta(t - t_j)$ represents the Dirac delta function for a spike arriving at time $t_j$, and $\omega$ is the synaptic weight that scales the amplitude of the incoming pulse.

\subsection{Experimental setup, datasets and energy estimation}

The proposed method is validated on a three-phase inverter platform (Fig.~\ref{fig:experimental_workflow}a) with parameters listed in Tables~\ref{tab:3} and \ref{tab:un_para}. Phase currents are measured by current sensors, sampled by the analog-to-digital converter (ADC) of a TI F28379D LaunchPad, and streamed to a host PC through a serial interface for preprocessing.

Experiments are conducted across all 22 inverter operational modes. For each loading condition, 1,000 current cycles are collected for each mode, resulting in 22,000 trajectory matrices. To improve robustness during mode transitions and current step changes, an additional transient dataset is included during training. For each transition, 167 transient trajectory matrices are generated from one fundamental cycle before and after the transition under a 5~kHz sampling frequency. These transient trajectory matrices are labeled as healthy, delaying the diagnosis decision until the trajectory matrix contains steady-state three-phase currents.

The method is evaluated under balanced loading, extremely unbalanced loading with 88\% phase current unbalance rate (PCUR), current amplitude step changes, and injected white Gaussian noise. For noise validation, 200 cycles per mode are sampled at each signal-to-noise ratio (SNR), producing 4,400 test matrices per SNR per loading condition. The CNN is trained on an NVIDIA RTX 4060 laptop GPU, and the trained parameters are transferred to NengoLoihi for SNN conversion and offline inference. The complete offline workflow is summarized in Fig.~\ref{fig:experimental_workflow}b.

\begin{table}[!ht]
\footnotesize
\centering
\caption{Key parameters for balanced loading}
\begin{tabular}{l c}
\toprule
\textbf{Parameter} & \textbf{Balanced System} \\
\midrule
Rated current            & \(2~\text{A}\)    \\
Switching frequency      & \(50~\text{kHz}\) \\
Sampling rate            & \(5~\text{kHz}\)  \\
Phase A/B/C R             & \(2.5~\Omega/2.5~\Omega/2.5~\Omega\)    \\
Inductance               & \(2~\text{mH}\)  \\
\bottomrule
\end{tabular}

\label{tab:3}
\end{table}

\begin{table}[!ht]
\footnotesize
\centering
\caption{Key parameters for unbalanced loading}
\begin{tabular}{l c}
\toprule
\textbf{Parameter} & \textbf{Unbalanced System} \\
\midrule
Rated current            & \(0.6~\text{A}\)    \\
Switching frequency      & \(50~\text{kHz}\) \\
Sampling rate            & \(5~\text{kHz}\)  \\
Phase A/B/C R             & \(8~\Omega/8~\Omega/100~\Omega\)    \\
Inductance               & \(2~\text{mH}\)  \\
\bottomrule
\end{tabular}

\label{tab:un_para}
\end{table}

Performance is evaluated based on diagnostic accuracy and energy per inference. For the CNN baseline, power consumption is profiled using the \texttt{nvidia-smi} utility with a 100~ms logging interval; idle power is subtracted to isolate the net inference energy. For the SNN, energy metrics are obtained via the Nengo \texttt{energy.summary} tool, which accounts for the dynamic power consumption of the neuromorphic cores during spike processing. Both SNN and CNN adopt the same architecture utilized in \cite{lei2026online}. Parameters are listed in \cref{tab:cnn_architecture}. It should be noted that the training and test datasets, as well as the trained model parameters, are pre-loaded onto the hardware prior to energy measurement. Therefore, the reported power consumption does not include the energy cost associated with communication between hardware platforms. The measurements reflect only the inference-related energy consumption.

Fig.~\ref{fig:experimental_workflow}e,f illustrates the GPU power log during training and 10 rounds of offline validation of the CNN-based diagnosis method proposed in \cite{lei2026online}. The GPU idling power is first measured, and an average value of 3.6~W is subtracted from the recorded power before energy calculation and plotting, following the procedure in \cite{ostrau2022benchmarking}. The trapezoidal integration method is adopted to estimate energy consumption, as it assumes linear variation between consecutive power samples over time, enabling more accurate estimation \cite{ostrau2022benchmarking}. To improve measurement accuracy, the testing process is repeated 10 times, and the final testing energy is obtained by averaging the results. Since each individual test lasts only approximately 0.2~s, a 100~ms logging interval produces only about two power samples, which can introduce significant measurement error. By extending the total testing duration through repeated rounds, more logging points are obtained, resulting in improved estimation accuracy. During both training and testing, GPU power consumption remains relatively stable, ranging from 73~W to 79~W with only minor fluctuations. Based on this range, the power estimation error is bounded within approximately 8\%, with the worst-case scenario corresponding to an estimated constant power of 73~W versus an actual constant power of 79~W. The energy consumption of all GPU-based methods is estimated using the same procedure.

For the Loihi-based SNN implementation, a different estimation approach is adopted, as on-chip power monitoring is not available. The energy consumption is calculated based on the energy-per-spike model and spike count. According to \cite{davies2018loihi}, the energy required to generate a single spike is approximately 23.6~pJ, as characterized during chip design and early silicon validation. During SNN execution using the NengoLoihi framework, the total number of spikes is recorded and multiplied by the energy per spike to estimate the overall energy consumption. It should be noted that Loihi energy is estimated from recorded spike counts and characterized Loihi energy-per-spike values. Direct board-level Loihi power measurement is not available in the current setup; therefore, the reported Loihi energy represents estimated neuromorphic inference energy rather than full-system energy. Future work will include direct board-level measurement in an integrated embedded implementation.

\begin{table}[!ht]
    \centering
    \caption{CNN/SNN Architecture and Training Parameters}
    \label{tab:cnn_architecture}
\begin{tabular}{@{}p{3cm}p{4cm}@{}}
        \toprule
        \textbf{Layer} & \textbf{Description} \\
        \midrule
        Convolution layer 1 & Filter size: $3 \times 3$; stride: $1 \times 1$; \newline filters: 32; padding: same \\
        Convolution layer 2 & Filter size: $3 \times 3$; stride: $1 \times 1$; \newline filters: 64; padding: same \\
        Convolution layer 3 & Filter size: $3 \times 3$; stride: $1 \times 1$; \newline filters: 128; padding: same \\
        Learning rate & $3 \times 10^{-4}$ \\
        Firing-rate scaling factor (SNN only) & 330~Hz \\
        Mini-batch size & 16 \\
        Epochs & 10 \\
        \bottomrule
    \end{tabular}

\end{table}

\backmatter

\bmhead{Data availability}
The datasets generated and analyzed during the current study are available from the corresponding author upon reasonable request.

\bmhead{Code availability}
The scripts used in this study are available from the corresponding author upon reasonable request.

\bmhead{Acknowledgements}
This work was supported in part by the U.S. National Science Foundation (NSF) under Award No. 2339806.
\bmhead{Author contributions}
X.L. conceived the study, developed the methodology, performed the experiments, analyzed the data, and wrote the manuscript. F.W. designed and built the experimental platform and assisted with the experimental validation. Y.L. supervised the project, acquired funding, and revised the manuscript. All authors reviewed and approved the final manuscript.

\bmhead{Competing interests}
The authors declare no competing interests.

\end{document}